\documentclass[dvips,a4paper,12pt]{article}
\usepackage{amssymb,amsmath,newalg,graphicx}
\newcommand{\singlespacing}{\let\CS=\@currsize\renewcommand{\baselinestretch}{1.5}\tiny\CS}
\makeatletter \oddsidemargin  -.1in \evensidemargin -.1in \textwidth
16cm \topmargin -.1in \textheight 21cm

\newcommand{\doublespacing}{\let\CS=\@currsize\renewcommand{\baselinestretch}{1.35}\tiny\CS}

\newtheorem{theorem}{Theorem}[section]
\newtheorem{proposition}[theorem]{Proposition}
\newtheorem{corollary}{Corollary}[theorem]

\newenvironment{proof}{
  \noindent\textbf{Proof}\ }{\hspace*{\fill}
  \begin{math}\Box\end{math}\medskip}

\begin{document}

\title{Finding Cliques of a Graph using Prime Numbers}\author{Dhananjay D. Kulkarni$^{a}$, Shekhar Verma$^{a}$, Prashant$^{b}$
\thanks{Corresponding author: Dhananjay D. Kulkarni email: dhananjay@iiitm.ac.in}\\
$^{a}$Indian Institute of Information Technology and Management,\\
 Gwalior, India.\\
$^{b}$Indian Institute of IT and Management, India.}

\date{}
\maketitle

\begin{abstract}
This paper proposes a new algorithm for solving maximal cliques for
simple undirected graphs using the theory of prime numbers. A novel
approach using prime numbers is used to find cliques and ends with a
discussion of the algorithm.
\end{abstract}

\section{Introduction:}
Graph-theoretic clustering techniques find their application in
myriad of problems in information science. One such technique is
finding all the cliques (maximal complete subgraphs) of a given
graph. A first general algorithm which enumerates all cliques of the
graph was given by Bierstone[1]. The Bierstone algorithm attempts to
find the cliques of the current node and its neighboring nodes which
can be merged with the subgraphs already generated to give the
maximal sub graph of the graph. A correction of the Bierstone's
algorithm was proposed by Gordon D. Mulligan and D.G. Corneil in
1972[2]

In this paper we propose a new notation for graphs using prime
numbers[3]. This is followed by an algorithm to enumerate all the
cliques of a general graph. The paper is commenced by a discussion.

\section{Notation Used:}
In general we consider a graph given by $G(V,E)$. Each vertex $u \in
V$ is identified by a unique prime number denoted by $v_u$. Every
vertex $u$, has a weight, denoted by $w_u = \prod_{i \in N[u]} v_i$
where $N[i]$ is the closed set of all vertices adjacent to $i$[4].

\begin{theorem}[Fundamental Theorem of Arithmetic]
\label{thm_FTOA}
Every positive number greater than $1$ can be written as a product of prime numbers in only one way.
\end{theorem}

\begin{corollary}
\label{cor_prime_divisor}
If $p,q_1,q_2, \ldots , q_n$ are all primes and $p \vert q_1 q_2 \cdots q_n$ then $p = q_k$ for some $k$ where $1 \leq k \leq n$.
\end{corollary}

\begin{proposition}
\label{prop_adj}
If $v_i \vert w_j$, there exists an edge from $j$ to $i$.
\end{proposition}

\begin{proof}
$w_j = \prod_{u \in N[v]} v_u$. From \ref{cor_prime_divisor} and the fact that $v_i$ is prime, $v_i = v_u$ for some $u \Rightarrow i \in N[j]$.
\end{proof}

Consider $g = \gcd(w_i, w_j)$. Using \ref{thm_FTOA} we can factorise $g$ in unique primes common to both $w_i$ and $w_j$. These are values of vertexes common to both $i$ and $j$.

\begin{proposition}
A clique can be uniquely identified by the product of the values of its participating vertices.
\end{proposition}

\begin{proof}
A clique is identified by the vertices participating in it. The value of each vertex is a unique prime. Thus the product is also unique.
\end{proof}

\section{Algorithm:}

\subsection{Theme:}

A graph is represented by a list $Q$ of tuples $\{v_u, w_u\}$ where each tuple represents a vertex. A vertex participates in all the cliques of its induced subgraph. Hence any arbitrary vertex is chosen as \textbf{pivot} - $p$ and two graphs (lists) are generated. $LeftQ$ represents the induced subgraph of pivot. $RightQ$ represents the subgraph in which pivot doesnot participate in any clique.

Consider any arbitrary vertex $u$. If $u \in N[p] \Rightarrow v_p \vert w_u$:
\newline
\newline
\textit{Case 1: For every clique in which $u$ participates, pivot is
one of the vertex.} This implies that $N[u] \subseteq N[p]
\Rightarrow w_u \vert w_p$. Thus $u \in LeftQ$. In the induced
subgraph, the neighborhood of $u$ is $N[u] - p$. Hence $w_u =
w_u/v_p$.
\newline
\newline
\textit{Case 2: For some cliques in which $u$ participates, pivot is
one of the vertex.} This implies that $N[u] \nsubseteq N[p]$ but
$N[u] \cap N[p] \neq \phi$. Here, $u \in LeftQ \wedge u \in RightQ$.
Two vertices $u_L$ and $u_R$ are created for $LeftQ$ and $RightQ$
respectively where $v_{u_L} = v_{u_R} = v_u$. $N[u_L] = N[u] \cap
N[p] - p \Rightarrow w_{u_L} = \gcd(w_u, w_p)/v_p$. $N[u_R] = N[u] -
p $, all vertices that fall in case 1.
\newline
\newline
Else, $u \in RightQ$. The cliques of the graphs thus generated are
found out by recursion. The terminating condition is a single vertex
or a null graph. The pivot is added to all the cliques of the
induced subgraph ($LeftQ$).

\subsection{Merging of vertices:}

If two vertices have the same neighborhood, they participate in the same cliques.
Hence, such vertices can be logically considered as a single vertex. Thus, if there exist $u_1, u_2 \ldots u_n$ such that $w_{u_1} = w_{u_2} = \ldots = w_{u_n}$, they can be merged in one vertex $u_m$ such that $v_{u_m} = v_{u_1} \times v_{u_2} \times \ldots \times v_{u_n}$ and $w_{u_m} = w_{u_1}$. Logically speaking, $u_m$ represents a clique $\{u_1, u_2 \ldots u_n\}$. A normal case would demand $n$ recursions (one for every vertex) against a single recursion after merging. Thus, if given graph is a clique, the vertices will coalesce into a single vertex reducing the amount of computation drastically. \\
\newline
\newline

\subsection{Algorithm is given below:}

\begin{algorithm}{Find-Clique}{Q}
CliqueQ \= \NIL \\
\CALL{Sort-By-Weight}(Q) \\
\begin{IF}{\vert Q \vert =\ 0}
    \RETURN CliqueQ
\end{IF} \\
\begin{FOR}{i \= 1 \TO \vert Q \vert} \\
    \begin{WHILE}{w_{Q[i]} =\ w_{Q[i+1]}}
        v_{Q[i]} \= v_{Q[i]} \times v_{Q[i+1]} \\
        \CALL{Remove}(Q, Q[i+1])
    \end{WHILE}
\end{FOR} \\
\begin{IF}{\vert Q \vert =\ 1}
    \CALL{Insert}(CliqueQ, v_{Q[1]}) \\
    \RETURN CliqueQ
\end{IF} \\
p \= Q[1] \\
LeftQ, RightQ, PivotQ \= \NIL \\
\begin{FOR}{j \= 2 \TO \vert Q \vert}
    \begin{IF}{v_{p} \vert w_{Q[j]}}
        w_{Q[j]} \= w_{Q[j]}/v_{p} \\
        \begin{IF}{w_{Q[j]} \vert w_{p}}
            \CALL{Insert}(LeftQ, Q[j]) \\
            \CALL{Insert}(NewQ, Q[j])
        \ELSE
            n_L, n_R \= \NIL \\
            v_{n_L}, v_{n_R} \= v_{Q[j]} \\
            w_{n_L} \= \CALL{GCD}(w_{p}, w_{Q[j]}) \\
            w_{n_R} \= w_{Q[j]} \\
            \CALL{Insert}(LeftQ, n_L), \CALL{Insert}(RightQ, n_R)
        \end{IF}
    \ELSE
        \CALL{Insert}(RightQ, Q[j])
    \end{IF}
\end{FOR} \\
\begin{FOR}{i \= 1 \TO \vert RightQ \vert}
    \begin{FOR}{j \= 1 \TO \vert PivotQ \vert}
        \begin{IF}{v_{PivotQ[j]} \vert w_{RightQ[i]}}
            w_{RightQ[i]} \= w_{RightQ[i]}/v_{PivotQ[j]}
        \end{IF}
    \end{FOR}
\end{FOR} \\
LeftCliqueQ, RightCliqueQ \= \NIL \\
LeftCliqueQ \= \CALL{Find-Clique}(LeftQ) \\
RightCliqueQ \= \CALL{Find-Clique}(RightQ) \\
\begin{FOR}{i \= 1 \TO \vert LeftCliqueQ \vert}
    LeftCliqueQ[i] \= LeftCliqueQ[i] \times v_p
\end{FOR} \\
CliqueQ = LeftCliqueQ + RightCliqueQ \\
\RETURN CliqueQ
\end{algorithm}

\subsection{Explanation of the algorithm:}

Initially all the tuples are sorted by weight to facilitate
identification of vertices having same neighborhood if so then they
are merged. If the input graph was a single vertex or a clique, the
degree of graph reduces to one after merging. The algorithm
terminates here reporting the value of the vertex as the clique in
the graph.

If the given graph is not a clique, then the first element is chosen
as the pivot. Based on this pivot, the graphs $LeftQ$ and $RightQ$
are generated. All the vertices belonging to \textit{case 1} are
also stored in $PivotQ$. These are then eliminated one-by-one from
the neighborhood of the vertices in $RightQ$. The cliques of
subgraphs $LeftQ$ and $RightQ$ are found out using recursion. The
pivot is then added to every clique of $LeftQ$. The two lists of
cliques are merged and the algorithm terminates.

\subsection{Analysis:}

The best case is when the given graph $G(V, E)$ is a clique
$C_{\vert V \vert}$. Then all the vertices merge into a single
vertex. The time required to find the clique then depends only upon
the time required to sort the vertices on the basis of their weight.
Using quick sort, best case complexity converges to $O(n \log(n))$.

The choice of the pivot governs the two subgraphs generated. Hence
if the \textit{pivot} is chosen such that that we have the induced
subgraph of order $\vert V \vert - 2$ and the other subgraph of
order $\vert V \vert - 1$. Such a case will occur only when the
\textit{pivot} participates in a clique with every other vertex
except one, say $y$ and $y$ participates with every other vertex
except \textit{pivot}. This gives,

\begin{equation*}
T(n) = T(n-2)+T(n-1)=O(2^n).
\end{equation*}

Thus the worst case complexity converges to $O(n)$.

\section{Conclusion:}

The theory of prime numbers can be used to perform simple set
operations like union and intersection using arithmetic functions
like gcd (greatest common divisor) and lcm(least common multiple).
The major drawback of this method is its storage space, while it
scores on simplicity.

Applying this principle to the theory of graphs, we can compute all
the maximal subsets of the graph efficiently by removing the
unwanted vertices by the process of merging, thus reducing the
complexity for some special type of graphs to $O(n \log(n))$.

More work however remains to be done. This spans the complexity in the average case, and also
possibility of a heuristic on choosing the $pivot$ such that the generated subgraphs
are more or less mutually exclusive.
\newline
\newline
\textbf{Acknowledgements:} We here by dedicate this work to Almighty
in full faith. We also acknowledge Dr. Shekhar Verma for his
guidance and support all through the research. Acknowledgements are
also due to Indian Institute of IT and Management, Gwalior, India
for extending all facilities and support in the completion of this
work.

\section{References:}
[1] J. Gary Auguston and Jack Minker \textbf{An Analysis of Some
Graph Theoretical Cluster Techniques}, Journal of the Association
for Computing Machinery, Vol. 17, No.5 October 1970, pp.571-588

[2] Gordon D. Mulligan and D.G. Corneil \textbf{Corrections to
Bierstone's Algorithm for Generating Cliques}, Journal of the
Association for Computing Machinery, Vol. 19, No.2, April 1972,
pp.244-247

[3] David M. Burton \textbf{Elementary Number Theory}, Universal
Book Publishers, New Delhi 1997.

[4] Douglas B. West \textbf{Introduction to Graph Theory}, Prentice
Hall of India, New Delhi 2003.
\end{document}